\documentclass{appolb}
\usepackage{graphicx}
\usepackage{amsmath,amssymb}
\usepackage{hyperref}
\usepackage{cite}
\usepackage{enumitem}
\setlist{noitemsep,topsep=2pt,parsep=2pt,partopsep=2pt,leftmargin=*}

\newcommand{\kt}{k_{\rm t}}
\newcommand{\qt}{q_{\rm t}}
\newcommand{\MSbar}{\overline{\text{MS}}}
\newcommand{\as}{\alpha_s}

\let\originalleft\left
\let\originalright\right
\renewcommand{\left}{\mathopen{}\mathclose\bgroup\originalleft}
\renewcommand{\right}{\aftergroup\egroup\originalright}
\def\beq{\begin{equation}}  
\def\eeq{\end{equation}}
\def\({\left(}
\def\){\right)}
\def\[{\left[}
\def\]{\right]}

\begin{document}
\title{Recent developments in small-$x$ resummation%
\thanks{Presented at Diffraction and Low-x 2018}%
}
\author{Marco Bonvini
\address{INFN, Sezione di Roma 1, Piazzale Aldo Moro~5, 00185 Roma, Italy}
}
\maketitle
\begin{abstract}
There has been a revived interest in small-$x$ resummation in recent times.
The main motivation was its success in describing small-$x$ HERA data without the inclusion
of non-perturbative corrections.
In this contribution I will review the recent developments in the field.
\end{abstract}

\vspace{1em}

Let us consider an observable $\sigma$, e.g.\ a DIS structure function, within the context of the collinear QCD
factorization theorem. It can be written in general as
\beq\label{eq:coll}
\sigma(Q^2) = \sum_{i=g,q}\int_x^1 \frac{dz}{z} \, C_i\(z,\as(Q^2)\) f_i\(\frac{x}{z},Q^2\),
\eeq
where $C_i$ are perturbative coefficient functions and $f_i$ are parton distribution functions (PDFs)
satisfying the DGLAP evolution equation
\beq\label{eq:DGLAP}
\mu^2\frac{d}{d\mu^2} f_i(x,\mu^2) = \sum_{j=g,q} \int_x^1 \frac{dz}{z} P_{ij}\(z,\as(\mu^2)\) f_j\(\frac xz,\mu^2\),
\eeq
where $P_{ij}$ are splitting functions and the sums extend over all partons.
It is well known that perturbative quantities computed in QCD may contain logarithmic enhancements in some regions.
This is for instance the case of $\as^n\frac1x\log^k\frac1x$ terms,
which appear in both splitting and coefficient functions in the singlet sector,
and become large at small-$x$, spoiling the perturbativity of the $\as$ expansion.
Resumming the small-$x$ logarithms cures the instability of the fixed-order perturbative results.
Small-$x$ resummation is based on the interplay of the previous two equations with the
$\kt$ factorization theorem
\beq\label{eq:kt}
\sigma(Q^2) = \sum_{i=g,q}\int_x^1 \frac{dz}{z} \int_0^\infty d\kt^2 \, {\cal C}_i\(z,\kt^2,\as\) {\cal F}_i\(\frac{x}{z},\kt^2\),
\eeq
where ${\cal C}_i$ are coefficient functions with off-shell initial-state partons (with off-shellness given by $\kt^2$),
and ${\cal F}_i$ are unintegrated $\kt$-dependent PDFs.
In the small-$x$ limit, the unintegrated gluon PDF is related to the integrated integrated PDF by
\beq\label{eq:Ff}
{\cal F}_g(x,\kt^2) = {\cal R}\, \frac{d}{d\kt^2} x f_g(x,\kt^2),
\eeq
where ${\cal R}$ is a scheme-dependent function. In the variant of the $\MSbar$ scheme usually adopted
in small-$x$ resummation, denoted $Q_0\MSbar$ scheme, ${\cal R}=1$.
The unintegrated gluon PDF satisfies the BFKL evolution equation%
\beq\label{eq:BFKL}
-x\frac{d}{dx} {\cal F}_g(x,\kt^2) = \int_0^\infty \frac{d\qt^2}{\kt^2} {\cal K}\(\frac{\kt^2}{\qt^2},\as\) {\cal F}_g\(x,\qt^2\),
\eeq
where ${\cal K}$ is the BFKL kernel.
Using Eq.~\eqref{eq:Ff} to transalte the BFKL equation into an equation for the integrated PDF,
it is then possible to require consistency between its solution and that of the DGLAP evolution Eq.~\eqref{eq:DGLAP}
to find constraints between the splitting functions and the BFKL kernel, called duality relation,
that allow to resum the small-$x$ logarithms in splitting functions.
Practically, the procedure is more complicated due to the perturbative instability of the BFKL kernel,
that requires a number of operations to be performed before obtaining a perturbatively stable result.
On top of this, the resummation of a class of subleading contributions originating from the running
of the strong coupling turns out to be very important, as it changes the nature of the small-$x$
behaviour.
Resummation at next-to-leading logarithmic (NLL) level matched to fixed next-to-leading order (NLO)
has been achieved by various groups (see e.g.~\cite
{Ciafaloni:2007gf,Altarelli:2008aj,White:2006yh}).

\begin{figure}[t]
  \centering
  \includegraphics[width=0.49\textwidth,page=1 ]{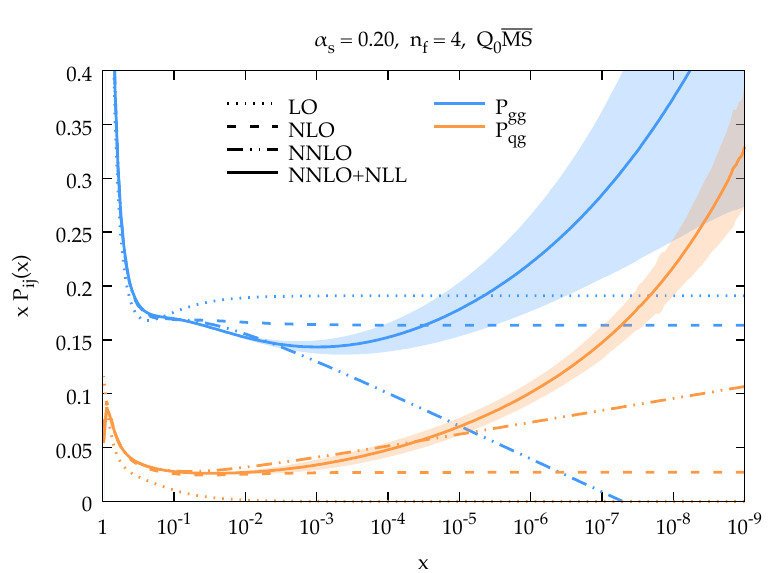}
  \includegraphics[width=0.49\textwidth,page=16]{images/plot_P_nf4_as020_mixed.pdf}
  \caption{Left: the $P_{gg}$ and $P_{qg}$ splitting functions at LO, NLO, NNLO and NNLO+NLL for $\as=0.2$, $n_f=4$.
    Right: N$^3$LO prediction for $P_{gg}$.}
  \label{fig:Pgg}
\end{figure}

In the recent Refs.~\cite{Bonvini:2016wki,Bonvini:2017ogt,Bonvini:2018xvt,Bonvini:2018iwt},
the formalism for small-$x$ resummation, in the approach of Altarelli-Ball-Forte (ABF), has been extended in many respects.
On top of (several) technical improvements, the main novelties that have been introduced are:
\begin{itemize}
\item matching the resummation to NNLO, to be able to construct DGLAP evolution at NNLO+NLL;
\item making a prediction of the (yet unknown~\cite{Davies:2016jie,Moch:2017uml}) N$^3$LO splitting functions at small $x$,
  and preparing all the ingredients to be able to match NLL resummation to N$^3$LO once available;
\item providing an uncertainty on resummed results from subleading logarithmic contributions;
\item releasing a public code, \texttt{HELL}~\cite{hell}, that implements the resummation and delivers resummed results for applications.
\end{itemize}
The first item of the list is particularly important, because the instability induced by small-$x$ logarithms
gets larger increasing the order. This is seen in Fig.~\ref{fig:Pgg} (left), where the $P_{gg}$ and $P_{qg}$ splitting functions
are shown at a low scale: the $\log\frac1x$ term at NNLO starts to grow for $x\lesssim10^{-2}$ and invalidate the perturbative expansion.
Once resummation is turned on, the behaviour changes substantially,
and the NNLO+NLL result deviates significantly from the fixed-order result.
A stronger effect is expected when matching the resummation to N$^3$LO,
because at this order extra powers of the log appear.
A prediction (based on the expansion of the resummed result) is shown in Fig.~\ref{fig:Pgg} (right).
The perturbative instability is apparent, especially when going to very small values of $x$.
However, subleading logarithmic contributions which cannot be fixed by NLL resummation
are potentially sizeable (difference between ``N$^3$LO approx'' and ``N$^3$LO asympt'' curves in the plot),
so this prediction carries a huge uncertainty and it may be useful only combined with other information
on the N$^3$LO result~\cite{Davies:2016jie,Moch:2017uml}.

\begin{figure}[t]
  \centering
  \includegraphics[width=0.49\textwidth]{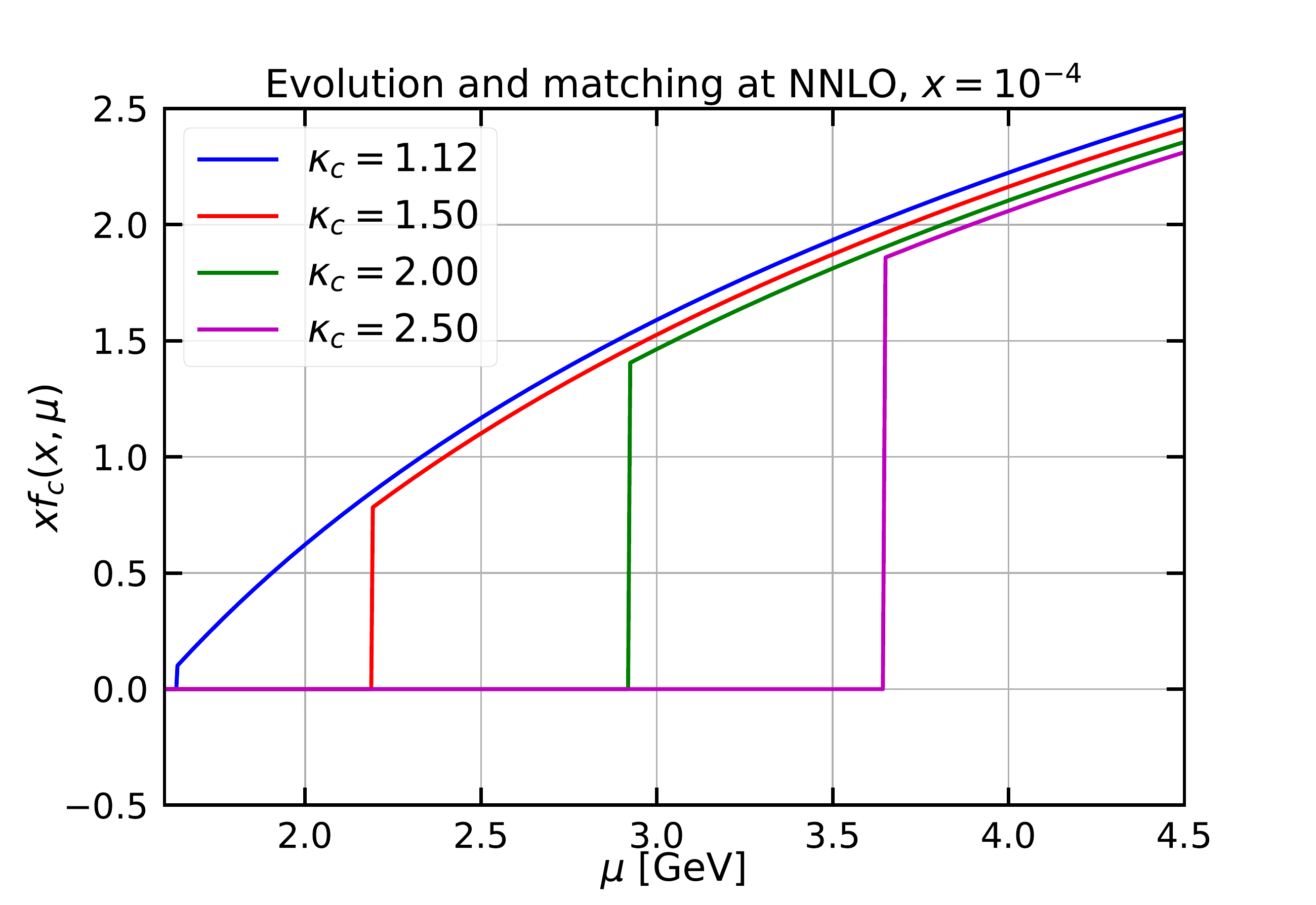}
  \includegraphics[width=0.49\textwidth]{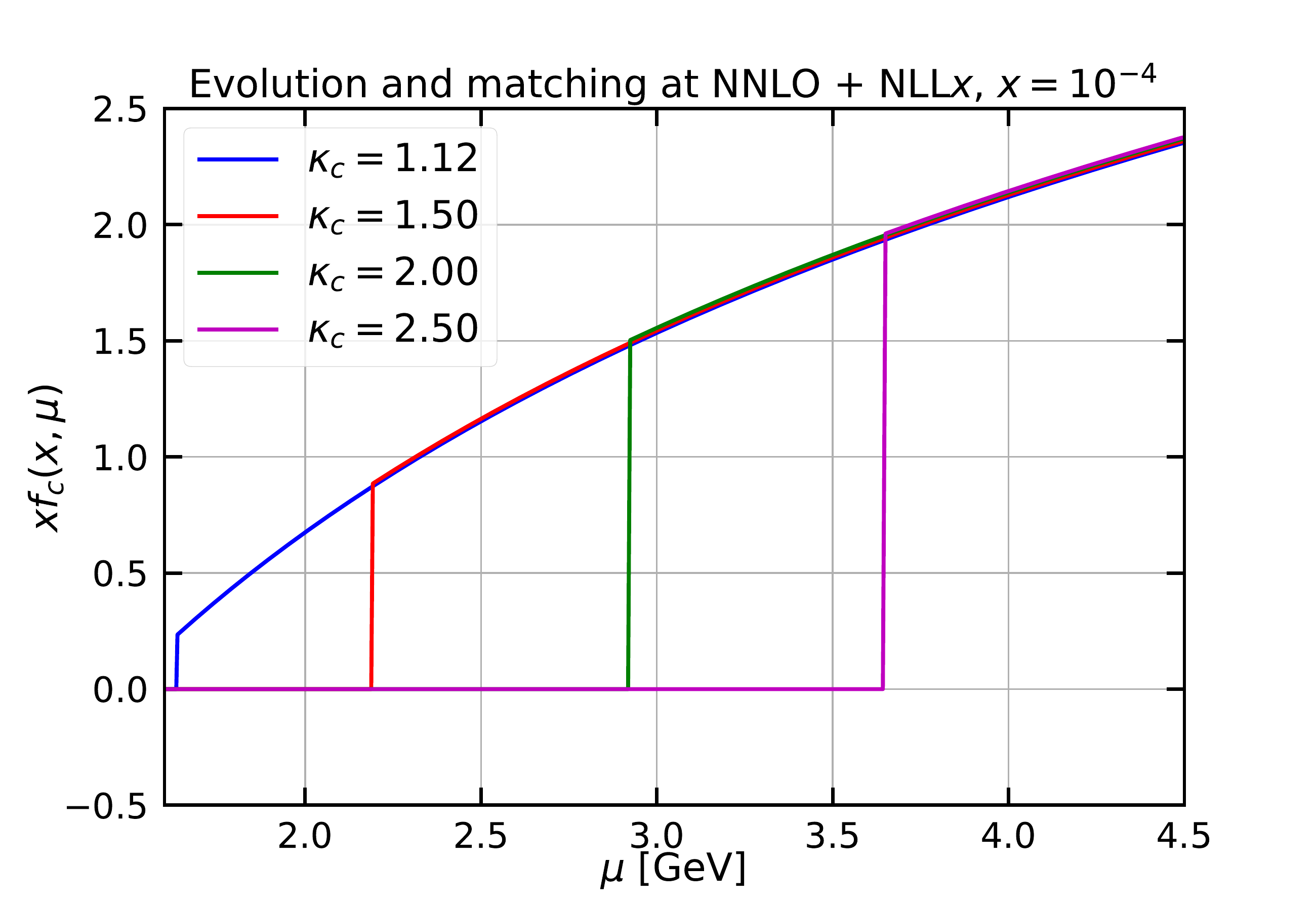}
  \caption{Fixed NNLO (left) and resummed NNLO+NLL (right) charm PDF generated perturbatively at different scales $\mu_c=\kappa_c m_c$,
    with $\kappa_c = 1.12,1.5,2,2.5$, at a small value of $x=10^{-4}$ (figure taken from Ref.~\cite{Abdolmaleki:2018jln}).}
  \label{fig:matching}
\end{figure}

The resummation of coefficient functions is based on the direct comparison of the collinear and $\kt$ factorization
formulae Eqs.~\eqref{eq:coll}, \eqref{eq:kt}, making use of a generalization of the relation Eq.~\eqref{eq:Ff}.
Moving to the Mellin $N$ space, and introducing the DGLAP evolution factors $U_{ij}(N,\mu^2,\mu_0^2)$ from a scale $\mu_0$ to a scale $\mu$,
we can rewrite Eq.~\eqref{eq:Ff} generalized to all flavours as
\beq
{\cal F}_k(N,\kt^2) = {\cal R}_{kj}\, \frac{d}{d\kt^2} U_{ji}(N,\kt^2,Q^2) f_i(N,Q^2),
\eeq
so that by comparison between the two factorization formulae we get
\beq
C_i\(N,\as(Q^2)\) =
\sum_{j=g,q} \int_0^\infty d\kt^2 \, {\cal C}_k\(N,\kt^2,\as\)\, {\cal R}_{kj}\,  \frac{d}{d\kt^2} U_{ji}(N,\kt^2,Q^2),
\eeq
which encodes the small-$x$ resummation provided the DGLAP evolution factors are themselves computed with resummed splitting functions.
This formulation of the resummation (introduced for the first time in Ref.~\cite{Bonvini:2016wki})
is equivalent to previous approaches~\cite{Catani:1994sq,Altarelli:2008aj,Ball:2007ra}, but it is very convenient from a numerical point of view,
and it allows for a simpler implementation of new processes in the resummation code \texttt{HELL}~\cite{hell}.
Also thanks to this new formulation, there have been a number of developments also in the context
of coefficient functions resummation~\cite{Bonvini:2017ogt,Bonvini:2018iwt}:
\begin{itemize}
\item resummation of all neutral- and charged-current DIS structure functions $F_2$, $F_L$ and $F_3$,
  both in the massless limit and including mass effects;
\item implementation of a variable flavour number scheme at small $x$ in $\MSbar$-like schemes;
\item resummation of heavy-quark matching conditions which give the initial conditions for the PDFs
when transitioning from a scheme with $n_f$ active flavours to a scheme with $n_f+1$ active flavours;
\item resummation of LHC observables (only Higgs production in gluon fusion so far, Drell-Yan is under investigation).
\end{itemize}
The third item turns out to be particularly interesting.
Indeed, the transition from the $n_f$ to the $n_f+1$ scheme happens at a (unphysical) matching scale,
that can be varied to assess the impact of unknown higher order contributions to the matching procedure.
Once resummation is included in the matching and in DGLAP evolution, the matching scale uncertainty
is drastically reduced at small $x$, thereby showing a stabilization of the perturbative expansion.
This is shown for the charm PDF in Fig.~\ref{fig:matching}. The gap between the various curves at large scale
(i.e.\ in the $n_f=4$ scheme) almost disappears once resummation is included.

Thanks to all these recent developments, and importantly to the availability of the public code \texttt{HELL}~\cite{hell}
that delivers resummed splitting and coefficient functions, it has been possible to perform two PDF fits including small-$x$ resummation,
one in the context of the NNPDF methodology~\cite{Ball:2017otu} and the other one using the xFitter toolkit~\cite{Abdolmaleki:2018jln}.
The striking effect of small-$x$ resummation is a dramatic improvement in the description of the low-$x$ low-$Q^2$ HERA data,
leading to a significantly different gluon (and quark-singlet) PDF at NNLO+NLL with respect to the NNLO fit at small $x$.

\begin{figure}[t]
  \centering
  \includegraphics[width=0.49\textwidth,page=1]{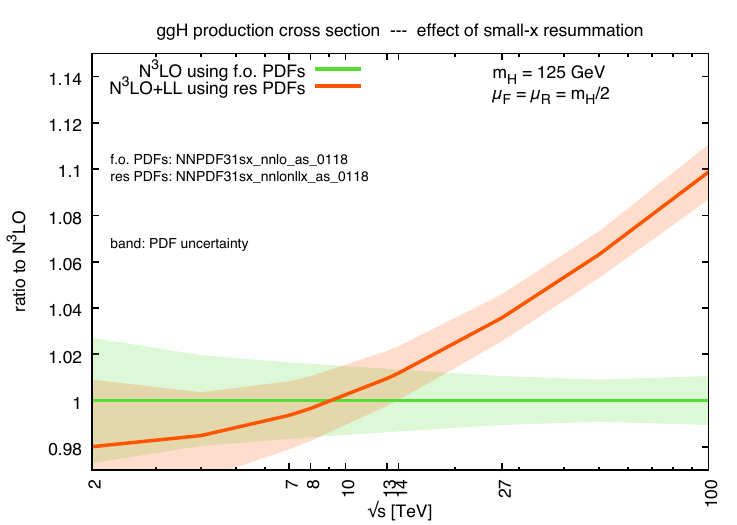}
  \includegraphics[width=0.49\textwidth,page=2]{images/ggH_Kfactor_summary3.pdf}
  \caption{Ratio of the resummed prediction of the Higgs cross section at N$^3$LO+LL to the fixed N$^3$LO result.
    In both plots the resummed PDF set is the one from Ref.~\cite{Ball:2017otu}, while the fixed order (NNLO) set
    used for the fixed-order prediction is either the baseline of the resummed fit~\cite{Ball:2017otu} (left plot),
    or the state-of-the-art NNPDF3.1 set of Ref.~\cite{Ball:2017nwa} (right plot).}
  \label{fig:ggH}
\end{figure}

To appreciate the importance of such effect, we show in Fig.~\ref{fig:ggH} the comparison of the fixed-order and resummed predictions
for the production of Higgs in gluon fusion at hadron colliders as a function of the collider energy~\cite{Bonvini:2018ixe,Bonvini:2018iwt}.
The effect of resummation (mostly coming from the use of resummed PDFs) is small and compatible within PDF uncertainty
with the fixed-order result up to approximately the current LHC energy.
For higher energies, the effect of resummation is a significant increase of the cross section,
rising with the energy, and reaching up to $\sim+10\%$ at a future circular collider of $100$~TeV.
This conclusion holds unchanged if using a different NNLO PDF set for the comparison,
for instance the state-of-the-art NNPDF3.1 set of Ref.~\cite{Ball:2017nwa} (right plot),
which has been fitted using a larger dataset.
Subleading logarithmic contributions may have sizeable effects~\cite{Bonvini:2018iwt}
and reduce (or enhance) the overall effect of resummation,
but the significance of the effect is likely independent of them.

\section*{Acknowledgments}
\noindent
This work is supported by the Marie Sk\l{}odowska-Curie grant HiPPiE@LHC, number 746159.

\addcontentsline{toc}{section}{References}
\bibliographystyle{jhep}
\bibliography{references}

\providecommand{\href}[2]{#2}\begingroup\raggedright\begin{thebibliography}{10}

\bibitem{Ciafaloni:2007gf}
M.~Ciafaloni, D.~Colferai, G.~Salam and A.~Stasto, \emph{{A Matrix formulation
  for small-$x$ singlet evolution}},
  \href{https://doi.org/10.1088/1126-6708/2007/08/046}{\emph{JHEP} {\bfseries
  0708} (2007) 046} [\href{https://arxiv.org/abs/0707.1453}{{\ttfamily
  0707.1453}}].

\bibitem{Altarelli:2008aj}
G.~Altarelli, R.~D. Ball and S.~Forte, \emph{{Small x Resummation with Quarks:
  Deep-Inelastic Scattering}},
  \href{https://doi.org/10.1016/j.nuclphysb.2008.03.003}{\emph{Nucl.Phys.}
  {\bfseries B799} (2008) 199}
  [\href{https://arxiv.org/abs/0802.0032}{{\ttfamily 0802.0032}}].

\bibitem{White:2006yh}
C.~D. White and R.~S. Thorne, \emph{{A Global Fit to Scattering Data with NLL
  BFKL Resummations}},
  \href{https://doi.org/10.1103/PhysRevD.75.034005}{\emph{Phys. Rev.}
  {\bfseries D75} (2007) 034005}
  [\href{https://arxiv.org/abs/hep-ph/0611204}{{\ttfamily hep-ph/0611204}}].

\bibitem{Bonvini:2016wki}
M.~Bonvini, S.~Marzani and T.~Peraro, \emph{{Small-$x$ resummation from HELL}},
  \href{https://doi.org/10.1140/epjc/s10052-016-4445-6}{\emph{Eur. Phys. J.}
  {\bfseries C76} (2016) 597}
  [\href{https://arxiv.org/abs/1607.02153}{{\ttfamily 1607.02153}}].

\bibitem{Bonvini:2017ogt}
M.~Bonvini, S.~Marzani and C.~Muselli, \emph{{Towards parton distribution
  functions with small-$x$ resummation: HELL 2.0}},
  \href{https://doi.org/10.1007/JHEP12(2017)117}{\emph{JHEP} {\bfseries 12}
  (2017) 117} [\href{https://arxiv.org/abs/1708.07510}{{\ttfamily
  1708.07510}}].

\bibitem{Bonvini:2018xvt}
M.~Bonvini and S.~Marzani, \emph{{Four-loop splitting functions at small $x$}},
  \href{https://doi.org/10.1007/JHEP06(2018)145}{\emph{JHEP} {\bfseries 06}
  (2018) 145} [\href{https://arxiv.org/abs/1805.06460}{{\ttfamily
  1805.06460}}].

\bibitem{Bonvini:2018iwt}
M.~Bonvini, \emph{{Small-$x$ phenomenology at the LHC and beyond: HELL 3.0 and
  the case of the Higgs cross section}},
  \href{https://doi.org/10.1140/epjc/s10052-018-6315-x}{\emph{Eur. Phys. J.}
  {\bfseries C78} (2018) 834}
  [\href{https://arxiv.org/abs/1805.08785}{{\ttfamily 1805.08785}}].

\bibitem{Davies:2016jie}
J.~Davies, A.~Vogt, B.~Ruijl, T.~Ueda and J.~A.~M. Vermaseren,
  \emph{{Large-$n_f$ contributions to the four-loop splitting functions in
  QCD}}, \href{https://doi.org/10.1016/j.nuclphysb.2016.12.012}{\emph{Nucl.
  Phys.} {\bfseries B915} (2017) 335}
  [\href{https://arxiv.org/abs/1610.07477}{{\ttfamily 1610.07477}}].

\bibitem{Moch:2017uml}
S.~Moch, B.~Ruijl, T.~Ueda, J.~A.~M. Vermaseren and A.~Vogt, \emph{{Four-Loop
  Non-Singlet Splitting Functions in the Planar Limit and Beyond}},
  \href{https://doi.org/10.1007/JHEP10(2017)041}{\emph{JHEP} {\bfseries 10}
  (2017) 041} [\href{https://arxiv.org/abs/1707.08315}{{\ttfamily
  1707.08315}}].

\bibitem{hell}
\href{https://www.ge.infn.it/~bonvini/hell/}{\texttt{https://www.ge.infn.it/$\sim$bonvini/hell/}}.

\bibitem{Catani:1994sq}
S.~Catani and F.~Hautmann, \emph{{High-energy factorization and small x deep
  inelastic scattering beyond leading order}},
  \href{https://doi.org/10.1016/0550-3213(94)90636-X}{\emph{Nucl.Phys.}
  {\bfseries B427} (1994) 475}
  [\href{https://arxiv.org/abs/hep-ph/9405388}{{\ttfamily hep-ph/9405388}}].

\bibitem{Ball:2007ra}
R.~D. Ball, \emph{{Resummation of Hadroproduction Cross-sections at High
  Energy}},
  \href{https://doi.org/10.1016/j.nuclphysb.2007.12.014}{\emph{Nucl.Phys.}
  {\bfseries B796} (2008) 137}
  [\href{https://arxiv.org/abs/0708.1277}{{\ttfamily 0708.1277}}].

\bibitem{Abdolmaleki:2018jln}
{\scshape xFitter Developers' Team} collaboration, H.~Abdolmaleki et~al.,
  \emph{{Impact of low-$x$ resummation on QCD analysis of HERA data}},
  \href{https://doi.org/10.1140/epjc/s10052-018-6090-8}{\emph{Eur. Phys. J.}
  {\bfseries C78} (2018) 621}
  [\href{https://arxiv.org/abs/1802.00064}{{\ttfamily 1802.00064}}].

\bibitem{Ball:2017otu}
R.~D. Ball, V.~Bertone, M.~Bonvini, S.~Marzani, J.~Rojo and L.~Rottoli,
  \emph{{Parton distributions with small-x resummation: evidence for BFKL
  dynamics in HERA data}},
  \href{https://doi.org/10.1140/epjc/s10052-018-5774-4}{\emph{Eur. Phys. J.}
  {\bfseries C78} (2018) 321}
  [\href{https://arxiv.org/abs/1710.05935}{{\ttfamily 1710.05935}}].

\bibitem{Ball:2017nwa}
{\scshape NNPDF} collaboration, R.~D. Ball et~al., \emph{{Parton distributions
  from high-precision collider data}},
  [\href{https://arxiv.org/abs/1706.00428}{{\ttfamily 1706.00428}}].

\bibitem{Bonvini:2018ixe}
M.~Bonvini and S.~Marzani, \emph{{Double resummation for Higgs production}},
  \href{https://doi.org/10.1103/PhysRevLett.120.202003}{\emph{Phys. Rev. Lett.}
  {\bfseries 120} (2018) 202003}
  [\href{https://arxiv.org/abs/1802.07758}{{\ttfamily 1802.07758}}].

\end{thebibliography}\endgroup

\end{document}